\documentclass[prl,twocolumn,showpacs,amsmath]{revtex4}
\usepackage[dvips]{graphicx}
\bibpunct{[}{]}{,$\!$}{n}{}{}

\newcommand{\eps}{\epsilon}
\newcommand{\V}[1]{\mathbf{#1}}
\newcommand{\Rm}[1]{\mathrm{#1}}
\newcommand{\Ol}[1]{\overline{#1}}

\newcommand{\eno}  [1]{\mbox{\!(\ref{#1})}}
\newcommand{\eqn}  [1]{\mbox{Eq.\! (\ref{#1})}}

\newcommand{\fig}  [1]{\mbox{FIG.~\ref{#1}}}

\newcommand{\figa} [2]{\mbox{FIG.~\ref{#1}#2}}

\begin{document}

\preprint{APS/123-QED}

\title{Evolutionary reconstruction of networks}

\author{Mads Ipsen}
\email{ipsen@fhi-berlin.mpg.de}
\author{Alexander S.~Mikhailov}
\affiliation{%
  Fritz-Haber-Institut der Max-Planck-Gesellschaft, Faradayweg 4-6,
  D-14195 Berlin, Germany
}%

\date{\today}

\begin{abstract}
Can a graph specifying the pattern of connections of a dynamical
network be reconstructed from statistical properties of a signal
generated by such a system? In this model study, we present an
evolutionary algorithm for reconstruction of graphs from their
Laplacian spectra. Through a stochastic process of mutations and
selection, evolving test networks converge to a reference
graph. Applying the method to several examples of random graphs,
clustered graphs, and small-world networks, we show that the proposed
stochastic evolution allows exact reconstruction of relatively small
networks and yields good approximations in the case of large sizes.
\end{abstract}

\maketitle

The operation of network-organized systems of different origins is
determined by the pattern of connections between their elements. The
principal framework for investigations of dynamical networks is
provided by graph theory~\cite{Diestel97,Bollobas85,Cv95,Ch97}.
Recently, properties of various social~\cite{Newman01c,WaSt98},
linguistic~\cite{CaSo01}, biochemical~\cite{Stange99} and neural
networks~\cite{WaSt98,Am00}, of the WWW~\cite{AlBa99,Kumar99} and the
Internet~\cite{Falout99}, have been analyzed.  Statistical mechanics
of systems with network organization has been reviewed \cite{AlBa01}.
Much effort is invested in the understanding how the structure of a
network is mapped to its function and determines its operation. On the
other hand, in applications ranging from bioengineering to
neurosciences one also needs to design networks with a given function
or reconstruct a network from its dynamics. Taking into account the
great complexity of network dynamics, explicit solutions of such
inverse problems of graph theory are difficult.  But graph
reconstruction may also be achieved without any knowledge of rules, by
running an artificial evolution process through which a network learns
to generate certain dynamics by adjusting its internal organization.
Indeed, evolutionary algorithms are known to yield efficient solutions
for complex optimization problems~\cite{Mi96}. For the problem of
graph reconstruction, such an approach has previously been
proposed~\cite{Mikh88,Mikh89}.

In this Letter, we present an evolutionary algorithm and apply it to
reconstruct graphs from their Laplacian spectra. Random graphs,
small-world networks and networks with cluster organization are
considered. We show that for relatively small graphs, exact
reconstruction within a reasonable evolution time is possible. For
larger graphs, the evolution leads to a network which provides a good
approximation of the target graph. Both the spectral properties as
well as other characteristic features of the reference network, such
as the diameter, clustering coefficient, and the average degree, are
well reproduced by the approximately reconstructed graph.

Any graph $G$ can be described by its adjacency matrix $\V{A}$ such
that $A_{ij} = 1$ if the nodes $i$ and $j$ are connected, and $A_{ij}
= 0$ otherwise.  A Laplacian spectrum of the graph $G$ is defined
\cite{Ch97} as the set of eigenvalues $\lambda_i$ of the matrix
$\V{T}$ with elements $T_{ij} = A_{ij} - m_i\delta_{ij}$ where
\mbox{$m_i = \sum_{j=1}^N A_{ij}$} is the degree of node $i$ and
$\delta_{ij}$ is the Kronecker symbol.

Laplacian spectra are closely related to dynamical properties of a
simple network: Consider a hypothetical linear ``molecule'' consisting
of $N$ identical particles connected by identical elastic strings. The
pattern of connections is defined by a graph $G$: a bond between
particles $i$ and $j$ in the network is present if the respective
element $A_{ij}$ in the adjacency matrix of the graph $G$ is equal to
unity and absent otherwise. This dynamical system is described by a
set of differential equations $\ddot{x}_i + \sum_{j=1}^N A_{ij} (x_i -
x_j) = 0$ for the coordinates $x_i$ of all particles. Obviously, the
vibration frequencies $\omega _k$ of such a molecule ($k = 0,1,\dots
,N-1$) are given by the eigenvalues $\lambda_k = -\omega_k^2$ of the
matrix $\V{T}$.  Note that one eigenvalue $\lambda_0$ always satisfy
$\lambda_0 = 0$ due to the translational invariance of this equation.
For this reason, the Laplacian spectra of a graph are also known as
the vibrational spectra~\cite{Ch97}.  Besides yielding a link to
dynamical networks, spectra provide a powerful invariant
characterization of graphs. Each graph of size $N$ is thus mapped into
a set of $N-1$ positive real numbers $\omega_i$. Various statistical
properties of graphs can be expressed or evaluated in terms of their
spectra \cite{Ch97}. Moreover, even though cospectral graphs
(topologically different graphs with the same spectra) are known to
exist, their fraction is very small~\footnote{For $N \leq 6$, there is
only one pair of connected cospectral graphs. For regular graphs, one
pair of size $12$, two pairs of size $16$, a cospectral quadruple of
size $28$, seven cospectral pairs of size $25$, and a set of 91
cospectral graphs of size $35$ have been found~(see~\cite{Cv95}).}.
Hence, with a high probability two graphs with coinciding spectra
would indeed be identical.

It is convenient to introduce the spectral density $\rho(\omega)$ for
a graph as a sum of narrow Lorentz distributions
\begin{equation}
  \rho(\omega) = C\sum_{k=1}^{N-1} 
  \frac{\gamma}{(\omega - \omega_k)^2 + \gamma^2} 
  \label{eq:SpectralDens}
\end{equation}
with the same width $\gamma$ and the normalization constant $C$ chosen
in such a way that $\int_0^{\infty }\rho (\omega )d\omega =1$. The
spectral distance $\eps$ between two graphs $G$ and $G_0$ with
densities $\rho(\omega)$ and $\rho_0(\omega)$ can then be defined as
\begin{equation}
  \eps =
  \sqrt{
    \int_0^\infty
    [ \rho(\omega) - \rho_0(\omega) ]^2d\omega}.
  \label{eq:SpectralDist}
\end{equation}

Our aim is to reconstruct graphs from their Laplacian spectra. Note
that the number $M_N$ of different graphs of a given size $N$ becomes
(super)astronomically large even for relatively small sizes. A lower
bound for $M_N$ is $2^{N(N-1)/2}/N!$, so even for $N = 50$ we have
$M_N > 1.9\times 10^{304}$. Therefore, finding an exact solution to
the inverse problem by subsequently testing all graphs is in practice
impossible.  Instead, we shall use an evolutionary procedure for graph
reconstruction.

Suppose that we want to reconstruct a certain reference graph $G_0$
with the spectral density $\rho_0(\omega )$. In order to do this, we
generate an arbitrary initial graph $G$ and introduce a stochastic
process of mutations and selection.  The mutations represent random
modifications of the pattern of connections whereas the selection is
based on the spectral distance~\eno{eq:SpectralDist} between two
graphs.

A \emph{mutation} of the graph $G$ first consists of deleting all
connections of a randomly chosen node $i$.  A new degree $m_i$ for
this particular node is then chosen at random between $1$ and $N-1$
followed by a random generation its $m_i$ new connections.  The
obtained mutated graph is denoted as $G'$.

To decide whether a mutation should be accepted (that is, to realize
\emph{a selection}), we calculate the spectral distance $\eps'$
between the modified graph $G'$ and the reference graph $G_0$.  This
is then compared with the spectral distance $\eps$ between $G$ and
$G_0$. If $\Delta \eps = \eps' - \eps <0$, the mutation is always
accepted. If $\Delta \eps >0$, the mutation is accepted with a certain
probability $p(\Delta \eps )=\exp (-\Delta \eps /\eps \theta )$. When
a mutation has been accepted, the graph $G$ is replaced by $G'$.

These two steps are applied iteratively and the evolution is continued
until the spectra are identical (\mbox{$\eps = 0$}) or the spectral
distance $\eps$ is smaller than a given threshold. Note that mutations
may be accepted even if $\Delta \eps >0$ to avoid that the evolution
gets trapped in a local minimum. The noise of the selection is
controlled by the ``temperature'' parameter $\theta $. The scheme is
similar to the Metropolis algorithm used in statistical mechanics and
complex combinatorial optimization~\cite{Me53,Kirk83}.

In the following, this procedure is applied to reconstruct three different
types of reference graphs.

\begin{figure}[t!]
  \centering
  \includegraphics*[width=\columnwidth]{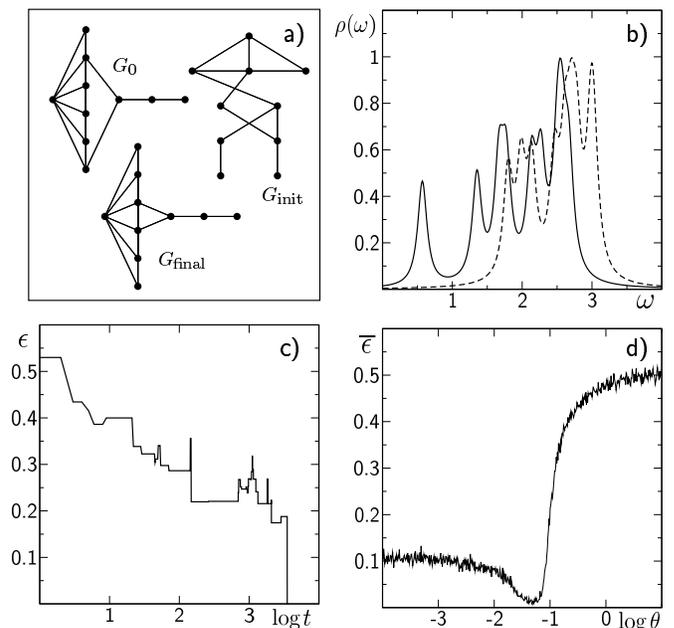}%
  \vspace*{-0.4cm}
  \caption{%
    Graphs $G_0$, $G_\Rm{init}$, and $G_\Rm{final}$ (note that $G_0$
    and $G_\Rm{final}$ are identical) (a). Spectral densities of $G_0$
    (solid line) and $G_\Rm{init}$ (dashed line) (b).  One stochastic
    evolution of the spectral distance $\eps(t)$ (c).  Dependence of
    the final mean spectral distance $\Ol{\eps}$ on the selection
    temperature $\theta$ (d).  Parameters are $\gamma = 0.08$ (b--d)
    and $\theta = 0.044$ (c).
  }%
  \label{fig:Simple}
\end{figure}

\emph{Random networks}. First we consider the case where the reference
network is a random graph of size $N$ and connection probability
$p$. As an example, we take a reference graph $G_0$ with $N = 10$ and
$p = 0.2$.  The initial graph $G_\Rm{init}$ is also random, but has a
higher connection probability $p = 0.9$. The two graphs and their
Laplacian spectra are shown in~\figa{fig:Simple}{a--b}. We then apply
the stochastic evolution, described above, with the selection
temperature $\theta = 0.044$ . The evolution of the spectral
difference $\eps$ is shown in~\figa{fig:Simple}{c}. The spectral
distance is gradually decreasing, with some fluctuations, until
eventually a transition occurs at $t\simeq 3500$, when the spectral
densities of the reference and test graph coincide. Examining the
final graph $G_\Rm{final}$ in~\figa{fig:Simple}{a}, we conclude that
it is indeed identical to the reference graph $G_0$. Note that though
the number of graphs of size $N = 10$ is of the order of $10^6$, the
exact reference graph has been reconstructed in only 3500 stochastic
iterations.

To determine the reliability of the reconstruction and its dependence
on the selection temperature $\theta$, a statistical investigation has
been performed. The variation of the mean spectral distance
$\Ol{\eps}$ as a function of $\theta$ is shown in~\figa{fig:Simple}{d}
where each point corresponds to the average over $10^3$ evolutions
each starting from a randomly chosen test graph with connection
probability chosen randomly from $[0,1]$. Each evolution was
terminated after $4 \times 10^4$ iterations.  We see that there is a
window of the selection temperature $\theta$ where fast convergence
takes place.  At the minimum $\theta = 0.04$ approx.\ 92 \% of all
evolutions converge exactly to the reference graph within the
specified time.

\begin{figure}[t!]
  \centering
  \includegraphics*[width=\columnwidth]{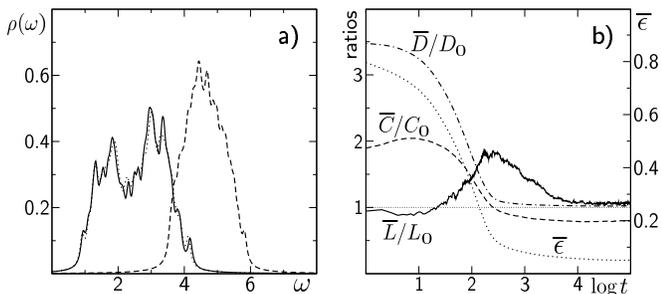}%
  \vspace*{-0.4cm}
  \caption{%
    Spectral densities of the clustered reference graph (solid line),
    and of the initial and final graphs (dashed and dotted lines) of
    one stochastic evolution (a).  Time dependences of the mean
    spectral distance $\Ol{\eps}(t)$ and of the ratios
    $\Ol{L}(t)/L_0$, $\Ol{C}(t)/C_0$, and $\Ol{D}(t)/D_0$ averaged
    over $10^3$ evolutions (b).  Parameters are $\gamma = 0.08$ and
    $\theta = 0.002$; for the reference graph $L_0 = 4$, $C_0 =
    0.263$, and $D_0 = 7.04$.
  }%
  \label{fig:Cluster}
\end{figure}

\emph{Clustered networks}. Next, we consider large clustered networks
representing a union of several random graphs with different
connection densities. As an example, a reference network $G_0$ of size
$N = 50$ with three clusters of high connection probability is
chosen. To prepare it, a sparse random graph of size $N$ with low
connection probability $p = 0.05$ is first generated. Then three
random dense clusters of size $N_\Rm{local} = 8$ with connection
probability $p_\Rm{local} = 0.8$ have been constructed and added to
the sparse graph.  The spectrum of $G_0$ exhibiting two distinct peaks
is shown by a solid line in \figa{fig:Cluster}{a}. The reconstruction
was tested by running $10^3$ stochastic evolutions starting from
different random graphs with connection probability chosen randomly
between $[0,1]$.  The spectra of one such initial graph and of the
corresponding final graph obtained after $10^5$ iterations are shown
in \figa{fig:Cluster}{a}.  Even though the exact reconstruction is not
reached in this case, the spectral densities of the final and of the
reference graphs in \figa{fig:Cluster}{a} are very similar ($\eps =
0.057$), and differ greatly from that of the initial graph.

Though exact reconstruction is not reached for the considered large
graph, statistical analysis reveals that the properties of the final
graph are close to that of the reference graph. \figa{fig:Cluster}{b}
shows the time dependence $\Ol{\eps}(t)$ of the mean spectral
difference averaged over the $10^3$ stochastic evolutions. We see that
$\Ol{\eps}$ decreases down to about 0.056 after $10^4$
iterations. Other important properties of the reference graph, such as
its diameter $L$, clustering coefficient $C$, and its mean degree $D$
(all defined as in the review \cite{AlBa01}), are also
well-reproduced.  This is seen from \figa{fig:Cluster}{b}, where the
time dependence of the mean ratios $\Ol{L}(t)/L_0$, $\Ol{C}(t)/C_0$,
and $\Ol{D}(t)/D_0$ is presented.

Similarity between graphs can also be discussed in terms of their
adjacency matrices: For any two graphs $G_1$ and $G_2$ with adjacency
matrices $\V{A}_1$ and $\V{A}_2$, a transformation $\V{F} =
\V{F}(\V{A}_1,\V{A}_2)$ can be introduced as
\begin{equation}
  \V{F} = 
  \V{U}_1^\Rm{T} \V{U}_2 \V{A}_2 \V{V}_2^\Rm{T} \V{V}_1.  
  \label{eq:SVD}
\end{equation}
Here the real matrices $\V{U}_{1,2}$ and $\V{V}_{1,2}$ are defined by
the singular-value decomposition~\cite{GolLoan96} of $\V{A}_1$ and
$\V{A}_2$.  If two graphs are identical and their adjacency matrices
only differ because of a different enumeration of nodes, the identity
$\V{A}_1 = \V{F}(\V{A}_1,\V{A}_2)$ holds and the difference
\mbox{$\V{\Delta} = \V{A}_1 - \V{F}$} is zero.  On the other hand, if
two graphs $G_1$ and $G_2$ do not coincide, the norm $\delta = 1/N
(\sum_{i,j}\V{\Delta}_{ij}^2 )^{1/2}$ of this difference can be used
as a measure of the distance between the graphs.

\begin{figure}[t!]
  \centering
  \includegraphics*[width=\columnwidth]{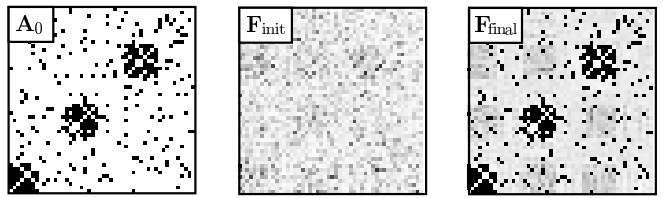}%
  \vspace*{-0.2cm}
  \caption{%
    Density plots of the adjacency matrix $\V{A}_0$ of the clustered
    reference graph and of the matrices $\V{F}_\Rm{init}$ and
    $\V{F}_\Rm{final}$ (see the text).
    }%
  \label{fig:CLDensity}
\end{figure}

In \fig{fig:CLDensity}, we visually display the matrices $\V{A}_0$,
\mbox{$\V{F}_\Rm{init} = \V{F}(\V{A}_0,\V{A}_\Rm{init})$}, and
$\V{F}_\Rm{final} = \V{F}(\V{A}_0, \V{A}_\Rm{final})$ where $\V{A}_0$,
$\V{A}_\Rm{init}$ and $\V{A}_\Rm{final}$ are the adjacency matrices of
the graphs $G_0$, $G_\Rm{init}$ and $G_\Rm{final}$.  Here the elements
in the matrices are represented by a square array of pixels using
gray-scale color maps whose limits are determined by the minimum and
maximum values of the respective matrix elements.  Even though the
matrix $\V{F}_\Rm{final}$ does not coincide with $\V{A}_0$, it is
already very close to it. The respective distances $\delta$ for the
initial and final graphs are $\delta_\Rm{init} = 0.41$ and
$\delta_\Rm{final} = 0.04$.

\begin{figure}[t!]
  \centering
  \includegraphics*[width=\columnwidth]{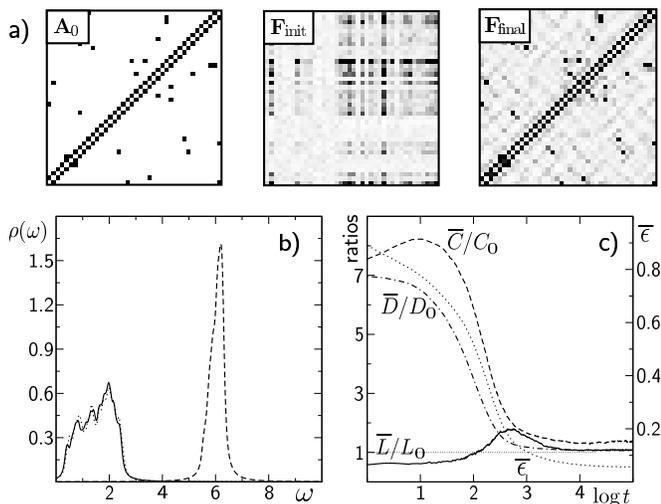}%
  \vspace*{-0.4cm}
  \caption{%
    Density plots of the adjacency matrix $\V{A}_0$ of a small-world
    reference graph and of the matrices $\V{F}_\Rm{init}$ and
    $\V{F}_\Rm{final}$ (a).  The corresponding spectral densities for
    the reference, initial, and final graphs (solid, dashed, and
    dotted lines, respectively) (b).  Time dependences of the mean
    spectral distance $ \Ol{\eps}(t)$ and of the ratios
    $\Ol{L}(t)/L_0$, $\Ol{C}(t)/C_0$, and $\Ol{D}(t)/D_0$ averaged
    over $10^3$ evolutions (c).  Parameters are $\gamma = 0.08$ and
    $\theta = 0.021$; for the reference graph $L_0 = 8$, $C_0 =
    0.063$, $D_0 = 2.65$.
    }%
  \label{fig:SmallWorld}
\end{figure}

\emph{Small~worlds}.  Now we consider a small-world graph
(see~\cite{WaSt98}) consisting of $N = 40$ nodes organized on a ring
with each node connected to its two neighboring nodes. In addition,
each node $i$ is connected to a randomly chosen node $j$ in the
network (\mbox{$j \neq i \pm 1$}) with the probability $p = 0.1$. The
adjacency matrix of this reference graph is visually displayed in the
left frame of \figa{fig:SmallWorld}{a}. To test the reconstruction
efficiency for this graph, we have performed $10^3$ stochastic
evolutions, each starting from a random graph with connection
probability chosen randomly from the interval $[0,1]$. The spectra of
the reference, initial and final graphs of one particular evolution
are shown in~\figa{fig:SmallWorld}{b}; we see that even fine
structures of the reference spectrum have been reproduced.  The
structure of the final graph is also close to that of the reference
graph as seen from the last two frames in~\figa{fig:SmallWorld}{a}
where the matrices $\V{F}_\Rm{init}$ and $\V{F}_\Rm{final}$ defined
by~\eqn{eq:SVD} are displayed (the respective distances $\delta$ for
the initial and final graphs are $\delta_\Rm{init} = 0.92$ and
$\delta_\Rm{final} = 0.01$).

The time dependence of the mean spectral distance $\Ol{\eps} (t)$ and
the ratios $\Ol{L}(t)/L_0$, $\Ol{C}(t)/C_0$, and $\Ol{D}(t)/D_0$
averaged over $10^3$ evolutions is shown in \figa{fig:SmallWorld}{c}.
After a transient of $10^4$ iterations, $\Ol{\eps} = 0.058$ and all
three ratios have approached unity implying that these characteristic
properties of the reference network also are well-captured by the
approximate reconstruction process.

Our analysis has revealed that the proposed evolutionary algorithm
provides an efficient method for exact or approximate reconstruction
of graphs from their Laplacian spectra. Based on the spectral density
only, such important properties of the reference network as its
diameter, clustering coefficient, and average degree are well
reproduced.  Moreover, approximately reconstructed networks have
similar adjacency matrices determining the pattern of connections
between nodes. Unique evolutionary reconstruction is not possible for
cospectral graphs, but these are, however, extremely rare~\cite{Cv95}.
Our numerical study has been performed using a fixed selection
temperature $\theta$ in the evolutionary algorithm; more refined
algorithms employing a time dependent temperature (similar to the
method of simulating annealing~\cite{Kirk83}) can also be implemented.

The spectral density~\eno{eq:SpectralDens} of a graph may be
interpreted as representing a power spectrum of a certain stochastic
signal $z(t)$.  This signal can be, for instance, generated by
vibrations of a damped elastic ``molecule'' under action of external
white noise. Such a linear dynamical system would be described by the
equations \mbox{$\ddot{x}_i + \gamma\dot{x}_i + \sum_{j=1}^N A_{ij}
(x_i - x_j) = \xi_i(t)$}, where $\xi_i(t)$ are independent random
forces and $A_{ij}$ is the adjacency matrix of the graph.  We have
shown that the structure of this simple model system can be
reconstructed from its temporal signal $z(t) = \sum_i a_i x_i(t)$ with
randomly chosen weights $a_i$. Thus, we see that important network
information can be encoded into a stochastic signal and then
efficiently recovered from it.

In essence, the proposed stochastic evolution can be regarded as a
learning process through which a test network, by adjusting its
internal structure, learns to approximate the dynamics generated by a
different system.  Similar approaches can be applied to solve other
problems.  For instance, approximations of large clustered graphs by
graphs of a smaller size can be constructed, and networks generating
stochastic signals with prescribed power spectra can be designed.

The results of our model study put forward questions whether network
organization of complex \emph{nonlinear} dynamical systems with
deterministic chaos can be reconstructed from their power spectra in
an evolutionary learning process and whether, generally, a nonlinear
network may learn to generate given complex chaotic dynamics by
iteratively adjusting its pattern of connections. A practical solution
of these problems would be important for a variety of applications.

MI acknowledges financial support from the Alexander von
Humboldt-Stiftung, Germany, and from the Danish Technical Research
Council (grant nr.~9901488).

\end{document}